\documentclass[aps,10pt,twocolumn,tightenlines]{revtex4}
\usepackage{graphics}
\usepackage{graphicx}
\usepackage{pstricks}
\newcommand{\be}{\begin{eqnarray}}
\newcommand{\beq}{\begin{equation}}
\newcommand{\eeq}{\end{equation}}
\newcommand{\ee}{\end{eqnarray}}

\newcommand{\bmp}{\noindent\begin{minipage}{16cm}}
\newcommand{\emp}{\end{minipage}\vskip 7mm} 

\newcommand{\lsim} {\buildrel < \over {_\sim}}
\newcommand{\gsim} {\buildrel > \over {_\sim}}
\usepackage{bm}
\usepackage{amsmath}
\usepackage{amsfonts}
\usepackage{bbm}

\begin{document}
\title{Majorana Dark Matter Cross Sections with Nucleons at High Energies}

\author{Yu Seon Jeong\footnote{ysjeong@cskim.yonsei.ac.kr} and C. S. Kim\footnote{cskim@yonsei.ac.kr}}
\affiliation{Department of Physics and IPAP, Yonsei University, Seoul 120-749, Korea}

\author{Mary Hall Reno\footnote{mary-hall-reno@uiowa.edu}}
\affiliation{Department of Physics and Astronomy, University of Iowa, Iowa City, Iowa 52242}

\date{\today}

\begin{abstract}
\noindent Non-relativistic dark matter scattering with nucleons is constrained by direct detection experiments.
We use the XENON constraints on the spin-independent
and spin-dependent cross section for dark matter scattering with nucleons to constrain a hypothetical Majorana fermionic
dark matter particle's couplings to the Higgs boson and $Z$ boson. In the procedure we illustrate the change
in the dark matter nucleon cross section as one goes from non-relativistic, coherent scattering
to relativistic, incoherent scattering. While the $Z$ invisible decay width excludes directly
couplings of dark matter to ordinary matter, by introducing a light $Z'$ portal to the dark sector,
a relatively large dark matter nucleon cross section can be preserved even with accelerator experiment
constraints for dark matter with a mass of $\sim 10$ GeV.

\end{abstract}

\maketitle

\section{Introduction}

The particle nature of dark matter (DM) is one of the biggest open questions at the intersection of particle physics and cosmology.
There are many candidates for DM in a variety of models \cite{Jungman:1995df,Bertone:2004pz,Hooper:2007hp,Drees:2012ji}. Cosmology and
large scale structure constrain the DM density in the universe, relative to the critical density, to be
$\Omega_{DM}h^2=0.112\pm0.006$ \cite{Drees:2012ji}, in terms of the Hubble constant $h$ in units of 100km/(s$\cdot$Mpc), for cold,
non-baryonic DM. This has implications for the annihilation cross section and mass of the DM
\cite{Drees:2012ji,Steigman:1984ac,Jarosik:2010iu} . Indirect constraints on the annihilation cross section and the DM
cross section with nucleons come from, for example, the absence of neutrino signals from the solar core and Earth's core
in neutrino telescopes \cite{Wikstrom:2009kw,Erkoca:2009by,IceCube:2011aj}. Finally, there
are constraints on the DM--nucleon cross section from direct detection experiments  \cite{Gaitskell:2004gd,Ahmed:2009zw,Ahmed:2010wy,
Armengaud:2011cy,Aprile:2011hi,Xenon10,Bernabei:2010mq,cogent,picasso,coupp,kims}.

The direct detection limits on the DM--nucleon cross section at low energy come primarily from experiments which measure the nuclear recoil in
elastic scattering of DM with nuclei \cite{Gaitskell:2004gd}. The DM masses typically covered by these
experiments \cite{Ahmed:2009zw,Ahmed:2010wy,Armengaud:2011cy,Aprile:2011hi,Xenon10,Bernabei:2010mq,cogent,picasso,coupp,kims}
are on the order of a few GeV through on the order of 1 TeV.
The momentum transfer is quite small, since the
weakly interacting massive particles are non-relativistic, so the scattering off of nuclei is coherent.
The energy dependence of the nuclear recoil depends on the sum of a term that is independent of the nuclear spin and another
term that depends on the net spin of the nucleus.
Depending on the particle type, the model of interactions and the nuclear target, the
cross section may be dominated by spin-dependent (SD) scattering or spin-independent (SI) scattering.

Because of the coherence enhancement
factor of $A^2$ for $A$ nucleons in the nucleus, the SI cross section limits are typically stronger than the SD
cross section limits
where there are cancellations between spins of nucleons.
For example, the cross section limit per nucleon for
SI couplings reaches a value of less than $\sigma_{SI}\sim 10^{-44}$ cm$^2$ for a DM mass $\sim 30$ GeV,
while for SD couplings, the limit is closer to $\sigma_{SD}\sim 5\times 10^{-39}$ cm$^2$ from direct detection experiments \cite{Drees:2012ji}.
With cross section limits, one can then determine constraints on coupling constants and masses in a given particle physics model.

In the low mass region, on the order of 10 GeV, the direct detection limits are inconsistent between
experiments, with  DAMA/LIBRA \cite{Bernabei:2010mq}
and CoGeNT \cite{cogent} reporting detection, and other groups reporting exclusion regions for the same parameters \cite{Ahmed:2009zw,Ahmed:2010wy,Aprile:2011hi}.
Even with this uncertainty, it is interesting to explore how the non-relativistic
$t$--channel scattering processes of direct detection experiments would translate to higher energy processes.

High energy DM may be produced in accelerator experiments and in cosmic ray interactions in
the atmosphere, for example. In accelerator experiments, typically the process of interest is
$s$--channel pair production of DM particles, here labeled by $\chi$.
In this letter, we consider the question of how direct detection constraints on the DM--nucleon cross section translate to the
high energy, deep-inelastic scattering (DIS) cross section for DM particles with nucleons,
a $t$--channel process. We use the
constraints on the  SI cross section \cite{Aprile:2011hi}
and the SD cross section
\cite{Xenon10}  limited by the XENON Collaboration
for the weakly interacting massive particle mass range of 10-1000 GeV.

The low energy cross section of direct detection experiments is non-relativistic and coherent while the high energy DIS cross section is relativistic and incoherent.
As a result, it is possible that effective operators at low energy that are poorly constrained can be the dominant interactions at high energy. If the DM couplings
are sufficiently large, then $\chi N$ interactions, in principle, could mimic neutral current (NC) interactions of neutrinos with nucleons.
The idea of DM  interactions giving the same signal as neutrino NC interactions has been discussed in Refs. \cite{ritz1} and \cite{ritz2}
for lower mass DM. Our goal here is to explore the possibility that while satisfying the direct detection constraints at low energies,
the DM $\chi N$ inelastic scattering cross section could be comparable to neutrino-nucleon cross sections at high energies for $m_\chi=10-1000$ GeV.

Specific particle physics models may predict either SI or SD dominated cross sections. General discussions of effective interactions
relevant to DM--nucleon interactions appear in, for example, Refs. \cite{Kurylov,Cheung:2010ua,Cheung:2012gi}.  There are results
on specific model assumptions, looking at the SI and SD constraints along with cosmological, collider physics  and neutrino
telescope constraints
in, e.g., Refs.  \cite{Cheung:2010ua,Cheung:2012gi,cpp,Barger:2010ng,foxetal,Rajaraman:2011wf,Goodman:2010ku}.

To avoid a multi-dimensional parameter space of a specific beyond-the-Standard Model, we consider here the simple
case of DM, a stable Majorana fermion. One example of Majorana fermionic DM is
discussed in Ref. \cite{Kouvaris:2007iq}. For simplicity, we treat the couplings of the Majorana fermion to the Higgs boson and $Z$ boson
as the only unknown parameters. This example will be sufficient to illustrate the translation of low energy, non-relativistic but coherent
scattering cross section limits  from the XENON collaboration \cite{Aprile:2011hi,Xenon10,Drees:2012ji}
to high energy, incoherent DIS of DM with nucleons.
We show that these limits are such that the DM--nucleon cross section may be comparable to the neutrino-nucleon cross section
for a range of energies and DM masses of order 10 GeV or 1 TeV.
However, the addition of LEP constraints and cosmological
constraints excludes the possibility of DM masquerading as neutrinos in the Standard Model (SM) NC interactions.
We discuss a possible low mass axial vector exchange that would permit an evasion of
these constraints, yet still allow neutrino-scale $\chi N$ interactions.

In the next section, we review the constraints of DM couplings with
the SM Higgs boson coming from the SI direct detection limit, and with the $Z$ boson from the SD direct detection limit, for DM
masses in the range of 10-1000 GeV. We show that $Z$ exchange for $m_\chi=10$ GeV, at the current upper limit for the $Z-\chi\chi$ coupling
from $\sigma(\chi N)$ direct detection limits, gives  NC cross section larger than the $\nu N$ NC cross section.
As long as the annihilation channel  to a real or virtual $Z$ is permitted ($\chi \chi \to Z$), the direct detection limits are not the
most constraining. For $m_\chi<M_Z/2$, the invisible decay width of the $Z$ provides the strongest constraint. For a range of masses,
if $\chi$ is the DM responsible for the bulk of the DM in the universe, the DIS cross section for
$\chi-N$ scattering will be considerably smaller than the neutrino-nucleon NC cross section.
An alternative scheme for DM couplings to ordinary matter through a light $Z'$ could allow DM DIS to
be roughly comparable to neutrino DIS, as discussed in Sec. III.

\section{Majorana fermionic DM and high energy DIS cross sections}

Majorana fermions are a useful example of particles in which the low energy direct detection constraints nevertheless allow a fairly large weak
interaction cross section for high energy DM DIS with nucleons. As discussed in Ref. \cite{Kurylov},
there are only two surviving terms in the effective interactions of Majorana DM with nucleons $N$ in the non-relativistic limit:
\be
{\cal L}\sim A \bar{\chi}\gamma^\mu \gamma_5 \chi \bar{N}\gamma_\mu \gamma_5 N + B\bar{\chi}\chi\bar{N}N\ .
\ee
We consider separately the couplings which
contribute to the SI and SD non-relativistic DM--nucleon cross sections, then evaluate the consequences
for deep inelastic DM--nucleon scattering. In Section III, we discuss other constraints on the DM couplings.

\subsection{Spin-independent effective operators}

If the DM is a Dirac fermion, a vector coupling with the $Z$ is permitted. The limit on the
$Z$ vector coupling to DM comes from the
SI cross section  and is of order $\sim 10^{-3}$ times the SM coupling.  If the $Z$ vector and axial vector couplings to a
Dirac fermionic DM particle are
related, the SI low energy cross section constraint would mean that high energy $\chi N$ scattering would be quite suppressed. It is
primarily for this reason that we are considering the DM as Majorana fermions.

For a Majorana fermion, the vector coupling with the $Z$ does not occur, so the
SI cross section limit on Majorana DM $\chi$ comes from scalar couplings.
There is an allowed effective interaction with quarks
\be
{\cal O}_f^H = c_f\bar{\chi}\chi \bar{f}f \ .
\ee
If the scalar exchange is a Higgs boson,
the quantity  $c_f$ depends on the Higgs boson mass, the nuclear matrix element of the $f=q$ operators and
the coupling of the DM to the Higgs boson. We proceed in our discussion with the assumption that the scalar is a Higgs boson and write
\be
{\cal L}^h=y_\chi \bar{\chi}\chi \, h\ .
\ee
The constraint on $y_\chi$ from the SI limit based on
the Higgs contribution to the SI cross section per nucleon, following Ref.
\cite{cpp}, is
\be
\label{eq:SI}
\sigma(\chi N\to \chi N)^{SI} \simeq 5\times 10^{-8} {\rm pb}\Biggl(\frac{y_\chi}{0.1}\Biggr)^2\Biggl(
\frac{115\ {\rm GeV}}{m_H}\Biggr)^4\ .
\ee
Using the XENON100 limits \cite{Aprile:2011hi,Drees:2012ji} and Eq. (\ref{eq:SI}), we set the upper values of $y_\chi$ to {0.25, 0.031, and 0.027} for  $m_\chi=10$, 25 and 100 GeV, respectively, for $m_H=125$ GeV. These values are listed in Table 1.

Given the DM coupling to the Higgs boson, at high energies, one contribution to the $\chi-N$ DIS cross section
comes from the effective operator
\be
\label{eq:ffh}
{\cal O}_f^H = \frac{y_\chi y_f}{Q^2+m_H^2}\bar{\chi}\chi \bar{f}f \ ,
\ee
for momentum transfer $Q$ and
$y_f=m_f/v$, where $f$ represents the quarks and $v=246$ GeV is the vacuum expectation value that characterizes the weak symmetry breaking scale.
For the DIS cross section, we
consider $E_\chi\geq 10 $ GeV, so $Q$ can be low, and the charm and $b$ components of the nucleon are not large at low $Q$.
There is an effective coupling of the Higgs boson to gluons
through a top quark loop \cite{Kniehl:1995tn}, however, the gluon contribution to $\chi N$ scattering does not change our
conclusions so we neglect it here.

\begin{table}[tb]
\begin{center}
\begin{tabular}{|c|c|c|r|}
\hline
$m_\chi$ [GeV] & $m_H$ [GeV]   & $y_\chi$  & $c_\chi^Z$ \\
\hline
10 & 115/125 &\  {0.21/0.25} & \ 1.77 \\			
\hline
25 & 125  &\  {0.031} & \ 0.40 \\				
\hline
100 & 125  &\  {0.027} & \ 0.56 \\				
\hline
\end{tabular}
\caption{\label{table:couplings} Approximate limits on the Higgs boson and $Z$ couplings to
DM based on direct detection limits on the SI \cite{Aprile:2011hi,Drees:2012ji}
and SD cross section for scattering with neutrons\cite{Xenon10}.}
\end{center}
\vspace{-0.6cm}
\end{table}

\subsection{Spin-dependent effective operators}

The
axial vector coupling of a Majorana particle to the $Z$ is allowed, where the coupling can be written as
\be
\label{eq:ffz}
{\cal L}_{M}^{ Z} = c_\chi^Z\frac{g}{2\cos\theta_W}\bar{\chi}\gamma_\mu\gamma_5 \chi Z^\mu \ .
\ee
Through an effective operator
$
O \sim (\bar{\chi}\gamma_\mu\gamma_5\chi )(\bar{f}\gamma^\mu \gamma_5 f ),$
the coupling $c_\chi^Z$ is constrained by the SD direct detection limits.
Cohen et al. in Ref. \cite{cpp} write, with the $Z$ exchange effective operator,  a low energy elastic cross section
which can be revised to account for just $\chi$ -- neutron ($n$) scattering:
\be
\sigma_{SD}(\chi n\to \chi n)\sim 3.2\times 10^{-4}\, {\rm pb}\Biggl(
\frac{c_\chi^Z}{0.1}\Biggr)^2 \ .
\ee
For $m_\chi=10$ GeV, the SD direct detection cross section limit \cite{Xenon10} is $\sim 0.1$ pb, while for $m_\chi=25$ GeV,
$\sigma_{SD}\lsim 5\times 10^{-3}$ pb and for $m_\chi=100$ GeV, $\sigma_{SD}\lsim 10^{-2}$ pb. These constraints
yield the values of $c_\chi^Z$
shown in Table \ref{table:couplings}.
{ We remark that the SD direct detection cross section limits on $\sigma_{SD}(\chi p)$ from, for example, PICASSO \cite{picasso}
and COUPP \cite{coupp}, translated to limits on $c_\chi^Z$ are weaker than the XENON10 limits on $\sigma_{SD}(\chi n)$ for
$m_\chi<1000$ GeV. The KIMS limits on $\sigma_{SD}(\chi p)$ \cite{kims} are competitive with the XENON10 limits at high mass (above 100 GeV). 
{For example, for $m_\chi=10^3$ GeV, the KIMS limit reduces $c_\chi^Z$ from the XENON10 limit by 25\%}.
We use the XENON10 limits here since the lower mass regime is of primary interest.}

\subsection{Deep inelastic scattering  cross section}

We now evaluate the deep inelastic cross section for Majorana fermion -- isoscalar nucleon ($N$) scattering for $E_\chi=10^3-10^{6}$ GeV.
The high energy cross section, at the parton level for elastic DM -- parton scattering in the $2\to 2$ process is
\be
\frac{d\hat{\sigma}}{dy}=\frac{1}{16\pi}\frac{\hat{s}-(m_\chi^2+m_f^2)}{\hat{s}^2-2\hat{s}(m_\chi^2+m_f^2)+(m_\chi^2-m_f^2)^2}
\overline{|{\cal M}|}^2 \ ,
\ee
where we label the parton mass with $m_f$. Here,
\be
y=\frac{p_1\cdot(k_1-k_2)}{p_1\cdot k_1}
\ee
for $2\rightarrow 2$ scattering, e.g., $\chi(k_1)+q(p_1)\to \chi(k_2)+q(p_2)$.

The high energy matrix element is model and process dependent.
For Higgs exchange with quarks with mass $m_f$, the spin-averaged matrix element squared is
\be
\overline{\mid {\cal M}\mid}^2 =  \Biggl(\frac{y_\chi m_f}{v^2(t-m_H^2)}\Biggr)^2(t-4 m_\chi^2)
(t-4 m_f^2)\ .
\ee
Using the couplings $y_\chi$ in Table I, we find that $\sigma (\chi N)/E_\chi$ from Higgs exchange, using the couplings in Table 1, 
{range between $2\times10^{-49}-3\times10^{-45}$ cm$^2$/GeV for $E_\chi=10^3$ GeV, to  $5\times10^{-47}-4\times10^{-45}$ cm$^2$/GeV for $E_\chi = 10^6$ GeV,}
where the larger cross sections correspond to $m_\chi=10$ GeV.
Variations of the Higgs mass, 115 $-$ 127 GeV, have little effect.

For $E_\nu=30-200$ GeV, the cross section with isoscalar nucleons for neutrinos is $(0.677\pm 0.014)\times 10^{-38}$ cm$^2$ and
for antineutrinos is $(0.334\pm 0.008)\times 10^{-38}$ cm$^2$. At higher energies, the neutrino cross section falls with energy
because of the effect of the propagator in the boson exchange \cite{Gandhi:1998ri,Jeong:2010za}. This same effect is seen
in the DM -- nucleon inelastic scattering cross section.
Thus,
Higgs scalar exchange is not an important process in which DM scattering with nucleons may mimic neutrino scattering at high energies.

The expression for Majorana fermion scattering with quarks via $Z$ exchange from Eq. (\ref{eq:ffz}) is
\begin{eqnarray}
\nonumber
\overline{\mid {\cal M}\mid}^2 &=& 2(c_\chi^Z)^2\Biggl( \frac{g^2}{4\cos^2\theta_W (t-m_Z^2)}\Biggr)^2\\
\nonumber & \times & \Biggl[
(c_V^2+c_A^2)\Bigl( (s-m_\chi^2-m_f^2)^2\\
\nonumber
&+& (u-m_\chi^2-m_f^2)^2-2t m_\chi^2\Bigr)\\
 &+& 2c_V^2 m_f^2 (t-4m_\chi^2)
-2 c_A^2 m_f^2 (t-8m_\chi^2)\Biggr] .
\label{eq:zexch}
\end{eqnarray}
In Eq. (\ref{eq:zexch}), $c_V=T_3-2Q\sin^2\theta_W$ and $c_A=T_3$, for $T_3=\pm1/2$ and electric charge $Q$ for the $Z$ couplings
to ordinary fermions.

\begin{figure}[tb]
\centering
\includegraphics[scale=0.35]{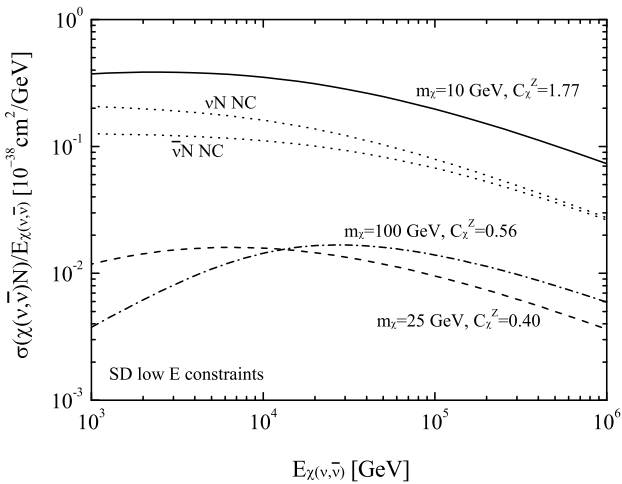}
   \caption{The inelastic cross section for $\chi$ scattering with nucleons including only $Z$ boson exchange from the SD limit on the
coupling $c_\chi^Z$. In this figure, the DM mass is set to 10, 25 and 100 GeV, and using $c_\chi^Z$ from Table \ref{table:couplings}. 
$\sigma(\nu N, \bar\nu N)$ NC cross sections as a function of $E_{\nu, \bar\nu}$ are also shown. }
\label{fig:zexchange}
\end{figure}

In contrast to Higgs exchange, based just on the low energy direct detection constraints on $c_\chi^Z$ for $ Z$ boson exchange in $\chi N$ interactions, the $\chi N$ NC cross section can be comparable to $\nu N$ NC interactions for low mass $m_\chi\lsim 10 $ GeV DM. 
Fig. \ref{fig:zexchange} shows the NC
cross section scaled by energy in units of $10^{-38}$ cm$^2$ for three DM masses and $m_\chi=10,\ 25$ and 100 GeV.
We use the CTEQ6.6 parton distribution functions \cite{cteq6} to evaluate the cross section.
This figure shows that from $E\sim\ 10^5$ GeV, the $\sigma(\chi N)\sim 2.5\times \sigma (\nu N)$. For higher masses, where the
low energy XENON SD cross section constraints are weaker, the $\chi N$ cross sections may be comparable to the neutrino -- nucleon
NC cross section.
At low energy, one sees kinematic effects due to the larger mass.

In the next section, we review other constraints on the $\chi \chi Z$ coupling which exclude this mass regime for Majorana DM which
couples directly to the $Z$. We discuss an alternative approach with a light $Z'$ exchange.

\begin{figure}[tb]
\centering
\includegraphics[scale=0.35]{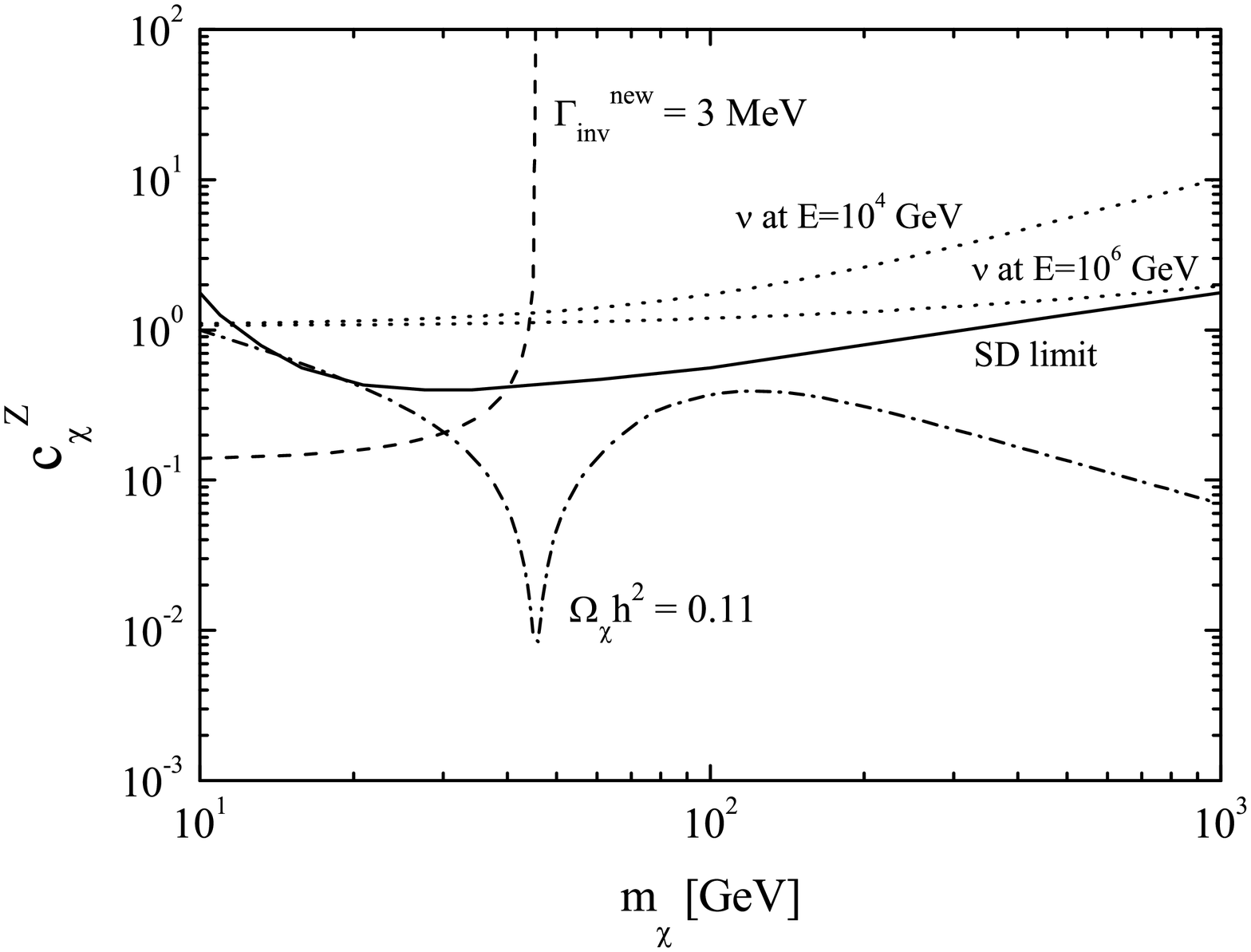}
   \caption{The coupling of DM to the $Z$ boson $c_\chi ^Z$ versus $m_\chi$. The dotted lines show the value of $c_\chi^Z$ such
that the $\chi N$ NC cross section equals $(\sigma(\nu N)+\sigma(\bar{\nu} N))_{NC}/2$ for $E=10^4$ and $10^6$ GeV, respectively.
The dashed line shows the upper bound on the coupling versus mass to be consistent with the limit on new physics
contributions to the invisible $Z$ decay width. 
The solid curve shows the the upper bound on the coupling translated from the XENON10 limit on the SD cross section with neutrons.
The dot-dashed line is the value of $c_\chi^Z$ such that the relic density of the DM $\Omega_{DM} h^2$ = 0.11 \cite{Kouvaris:2007iq}. 
}
\label{fig:zlimits}
\end{figure}

\section{Discussion}

Based on just the direct detection limits on the $\chi \chi Z$ coupling, one finds a mass region where, given high energy { Majorana}
DM, the
$\chi N$ cross section with $Z$ exchange is comparable to the neutrino -- nucleon NC cross section.  In Fig. 2, the dotted lines
show the value of $c_\chi^Z$ such
that the $\chi N$ NC cross section equals $(\sigma(\nu N)+\sigma(\bar{\nu} N))_{NC}/2$ for $E=10^4$  GeV and $E=10^6$ GeV, as a function of $m_\chi$.
The solid curve in Fig. 2 corresponds to the XENON10
limit on the SD low energy cross section with neutrons, translated to an upper bound on $c_\chi^Z$, as a function of $m_\chi$.

The SD direct detection limit is not the only constraint on $Z$ couplings to { Majorana} DM.
One indirect limit comes from assuming that $\chi$ makes up most of the DM in the
universe, shown in Fig. 2 with the dot-dashed line. This is the value of $c_\chi ^Z$ such that $\chi$  is the DM for which
$\Omega _{DM}h^2=0.11$ \cite{Kouvaris:2007iq}.  Particularly for $m_\chi\simeq 10$ GeV, the Majorana
DM $\chi$ may satisfy both this cosmological constraint, the SD limit on coupling, yet have a neutrino-like NC cross section at high energy.

The more stringent constraint on direct $\chi\chi Z$ couplings comes from $Z$ decays.
In Fig. 2, the dashed line labeled
$\Gamma_{\rm inv}^{\rm new}=3$ MeV shows the values of $c_\chi^Z$ versus $m_\chi$ such that the
effective contribution to the invisible decay width of the $Z$ is equivalent to an additional number of neutrinos
$\Delta N_\nu= 0.02$ \cite{pdg}. The mass range $m_\chi<M_Z/2$ is essentially excluded.
This excludes light mass ($\sim 10$ GeV) DM
that could mimic neutrino signals if its primary coupling to the SM particles is directly
via $Z$ exchange.

For the parameter space where $m_\chi>M_Z/2$ with $\chi$ the DM in the universe (the dot-dashed line), $c_\chi^Z$ lies below the SD limit,
and well below the region of parameter space where the $\chi N$ cross section
is comparable to the $\nu N$ cross section. For example, $c_\chi^Z=0.37$ for $m_\chi=100$ GeV if $\chi$ annihilation through the $Z$
is the process responsible for the DM density of the universe. For this parameter choice, $\sigma(\chi N)$ is about 1\% of the
average neutrino plus antineutrino NC cross section at $E=10^3$ GeV, growing to about 9\% of the average NC cross section at $E=10^6$ GeV.
For $m_\chi=1$ TeV, the allowed coupling to
the $Z$ from SD direct detection experiments is large enough that $\chi N$ scattering could be
comparable to $\nu N$ scattering for high energies, however, $\chi$ could not be the principle DM
in the universe because its annihilation cross section would be too large. Such large values of $c_\chi^Z$ for $m_\chi\gsim
50$ GeV is further constrained by neutrino induced muon events in IceCube \cite{IceCube:2011aj}.

With hints of a signal of a $\sim 10$ GeV DM particle coinciding with a prediction of the correct DM abundance
in the universe, we revisit the case of $m_\chi=10$ GeV.
The invisible $Z$ decay width  is the major constraint for
the simple extension of the SM with a low mass DM particle considered thus far.
A further extension of the SM that could yield neutrino-like $\chi N$ cross sections, and still be
accommodated by current experiments, is a light portal model. Portal models, first introduced in
Ref. \cite{holdom}, have a dark sector connected to the SM particles through mixing
of a new scalar or gauge boson with the SM Higgs boson or the SM gauge bosons.
While a Higgs portal model or vector boson portal model is more theoretically economical,
we will assume here
an axial vector portal, where a gauge boson $Z'$ with axial vector couplings with $\chi$ mixes with
the SM $Z$ with a small mixing angle denoted by $\xi$.

The $Z'$ couplings to the
SM particles are suppressed by a factor of $\xi$ in the amplitude, so $Z'$
contributions to SM $\to$ SM processes are suppressed by $\xi^4$. Processes with
SM $\to \chi \chi$ are suppressed by $\xi^2$. With a sufficiently small $\xi$, even
a relatively low mass $Z'$ is permitted.

As an example, we consider $m_\chi=10$ GeV with an axial vector portal connection to the SM particles.
We assume that the $Z'$ couplings to the SM particles are the same as $Z$ couplings, multiplied by $\xi$.
{}From Fig. 2, approximate scaling shows that if $c_\chi^Z=1.77$ from direct detection
constraints, in the portal case,
\be
c_\chi^{Z'}\xi M_Z^2/M_{Z'}^2 = 1.77\ .
\ee
To look at a specific example, we take $M_{Z'}=15 $ GeV to discuss experimental constraints.
With this mass, $c_{\chi}^{Z'}\xi \simeq 0.048$.

\begin{figure}[ht]
\centering
\includegraphics[scale=0.35]{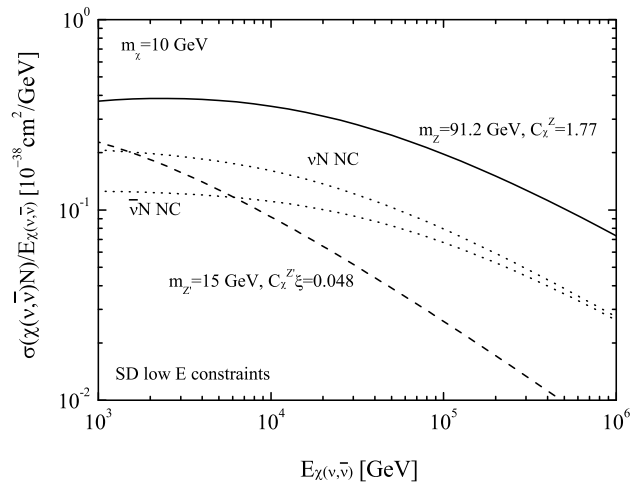}
   \caption{The inelastic cross section for $\chi$ scattering with nucleons including only $Z'$ boson exchange ($m_{Z'}=15$ GeV),
   represented by the dashed line, from the SD limit on the coupling $c_\chi^{Z'}\xi=0.048$.
In this figure, the DM mass is set to 10 GeV.
The solid and dotted curves, from Fig. 1, are shown for comparison.
}
\label{fig:zpexchange}
\end{figure}

Fig. 3 shows with a dashed line the $\chi N$ inelastic cross section for $c_\chi^{Z'}\xi=0.048$, $m_\chi = 10$ GeV and
$m_{Z'}=15$ GeV. The inelastic cross section is not as large as the inelastic cross section rescaled
for $Z$ exchange because the typical $Q$ is of order $Q\sim m_{Z'}$ at high energy. At a given value of Bjorken
$x$, the parton distribution functions are smaller for $Q\sim m_{Z'}$ than they are for $Q\sim m_{Z}$.
As the figure shows, the DIS cross section for $\chi N$ scattering is comparable to the $\nu N$ and
$\bar{\nu}N$ cross section only for $E_\chi \sim 10^3-10^4$ GeV.

Even with this fairly large DIS cross section, the other constraints for $m_\chi=10$ GeV are satisfied.
The effective coupling for low energy $\chi\chi$ annihilation relevant to $\Omega_{\chi} h^2$ are unchanged.
If we further assume that $c_\chi^{Z'}=1$,
the boson mixing angle is $\xi=0.048$.
This small mixing of $Z-Z'$ means that $Z\to \chi\chi$ is suppressed by $\xi^2\sim 2\cdot 10^{-3}$, so
$\Gamma_{\rm inv}^{\rm new}$ limits are satisfied. The factor of $\xi^4$ in SM $\to$ SM processes
ensures that even a low mass $Z'$ has suppressed contributions.

Finally, one should consider collider limits on production of $\chi\chi$. As discussed experimentally \cite{Chatrchyan,ATLAS,CDF,CMSpht,CMSDM,CMSjet}
and theoretically (e.g., in Refs. \cite{foxetal,Rajaraman:2011wf,Goodman:2010ku}), 
DM signatures at colliders may appear as mono-jets plus missing energy or mono-photons plus missing energy.
For $s$-channel production of a heavy boson, an effective four-fermion theory with partons and $\chi\chi$ can lead to
strong constraints on the scale of the boson mass. For the scenario presented here, with a light
mediator with $m_{Z'}<2 m_\chi$, resonant production of $\chi\chi$ is not possible
and high energies give cross sections suppressed by $1/\hat{s}$, as noted in Ref. \cite{foxetal}.

In conclusion, while we have not presented a specific model for Majorana DM coupling to ordinary
matter through an axial vector portal, we have shown that there is the potential for neutrino-like
couplings of the DM to nucleons. { We have focused on Majorana DM because the spin-dependent couplings dominate
the direct detection cross section. Scalar dark matter would have spin-independent Higgs exchange
at low energy that have strong SI constraints from direct detection \cite{Cheung:2012gi}.} For high energy
{ Majorana} DM, this may lead to events that are
indistinguishable from neutrino-nucleon NC events. Should a ratio of neutrino NC to charged current DIS rates be measured in experiments, the possibility of a high energy { Majorana}
DM flux mixed with the neutrino flux should be explored in the context of specific models.
\\

\noindent{\bf Acknowledgments}

We thank I. Sarcevic for discussions. This work was supported in part by Us Department of Energy contract  DE-FG02-91ER40664.
The work of C. S. K. and Y. S. J. was supported by the National Research Foundation of Korea (NRF) grant funded by the Korea government
of the Ministry of Education, Science and Technology (MEST) (No. 2011-0027275), (No. 2012-0005690) and (No. 2011-0020333).


\end{document}